\documentclass{article}

\usepackage{arxiv}
\usepackage{authblk}
\usepackage[utf8]{inputenc}
\usepackage[T1]{fontenc}
\usepackage{lineno}
\usepackage{longtable,makecell}
\usepackage{amsmath}
\usepackage{graphicx}
\usepackage{fancyhdr}
\usepackage{float}
\usepackage{textcomp}
\usepackage{indentfirst}
\usepackage[usenames,dvipsnames]{color}
\usepackage{url}
\usepackage{microtype}
\usepackage[ruled]{algorithm2e}
\usepackage{subfig}
\usepackage{mathrsfs}
\graphicspath{ {./}}

\title{Privacy-Preserving ECG Data Analysis with Differential Privacy: A Literature Review and A Case Study}

\author[1]{Arin Ghazarian}
\author[2]{Jianwei Zheng}
\author[3]{Cyril Rakovski}
\affil[1,2,3]{Schmid College of Science and Technology, Chapman University, Orange, California, USA}
\affil[1]{Corresponding author: Arin Ghazarian (Email: ghazarian@chapman.edu) }
\begin{document}
\maketitle
\begin{abstract}
Differential privacy has become the preeminent technique to protect the privacy of individuals in a database while allowing useful results from data analysis to be shared. Notably, it guarantees the amount of privacy loss in the worst-case scenario. Although many theoretical research papers have been published, practical real-life application of differential privacy demands estimating several important parameters without any clear solutions or guidelines. In the first part of the paper, we provide an overview of key concepts in differential privacy, followed by a literature review and discussion of its application to ECG analysis. In the second part of the paper, we explore how to implement differentially private query release on an arrhythmia database using a six-step process. We provide guidelines and discuss the related literature for all the steps involved, such as selection of the $\epsilon$ value, distribution of the total $\epsilon$ budget across the queries, and estimation of the sensitivity for the query functions. At the end, we discuss the shortcomings and challenges of applying differential privacy to ECG datasets.
\end{abstract}


\flushbottom

\thispagestyle{empty}

\section*{Introduction}
\begin{figure}
    \centering
	\includegraphics[scale =0.27]{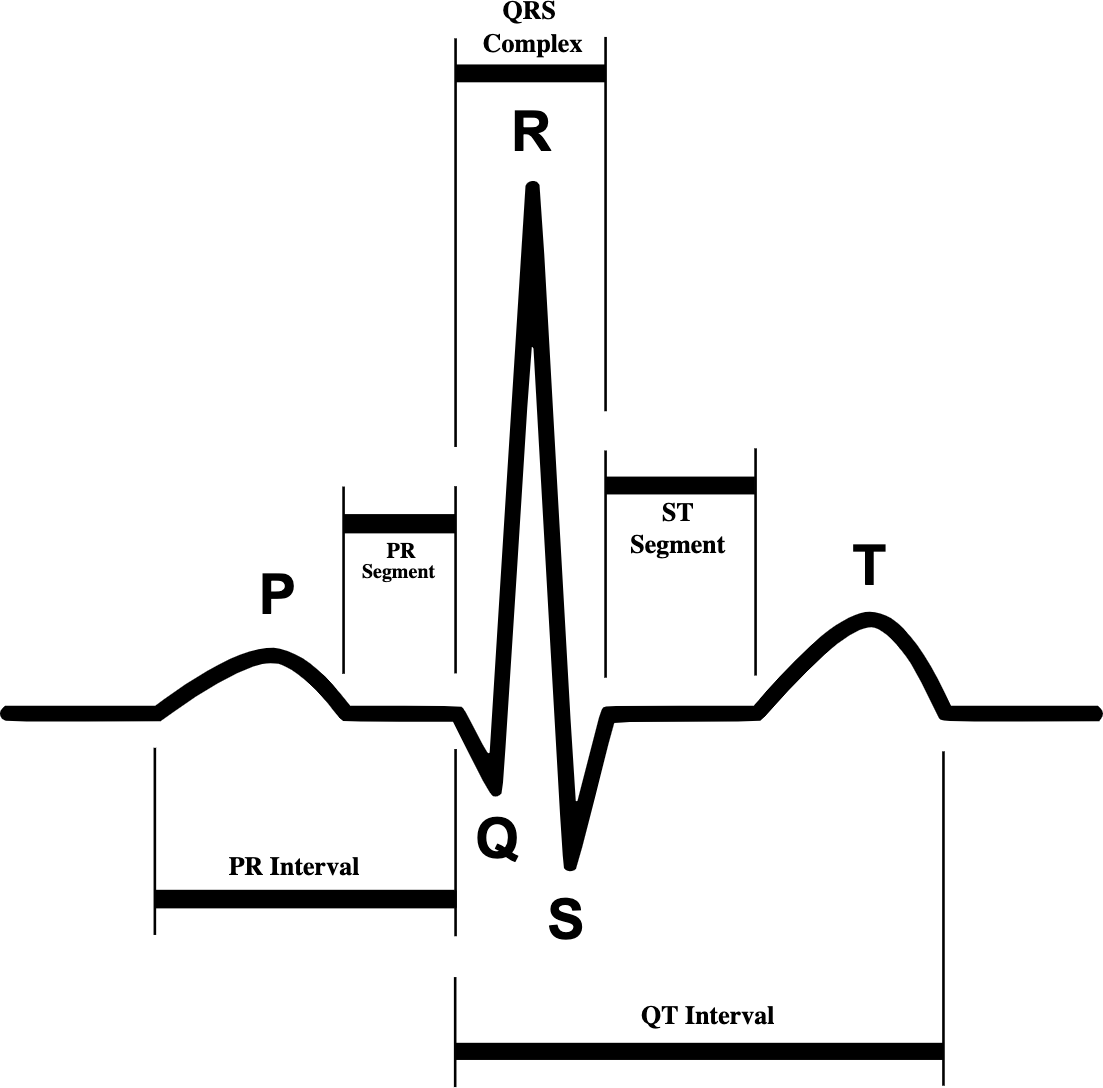}
	\caption{ECG waveform and segments in lead II for normal cardiac cycle}
	\label{fig1:ECG waves}
\end{figure}

\begin{figure}
	\centering
	\includegraphics[scale =0.25]{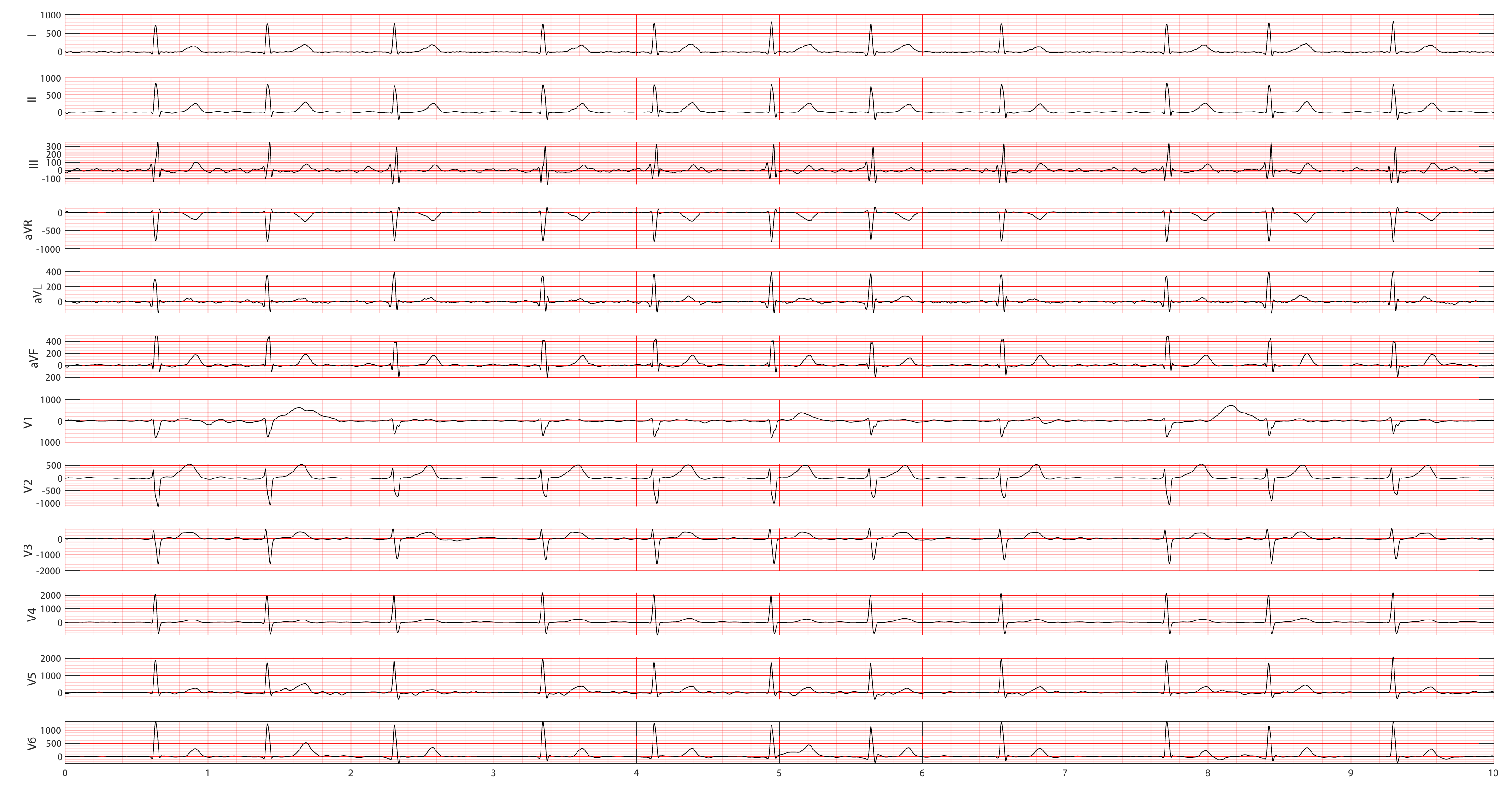}
	\caption{A twelve-lead ECG showing atrial fibrillation rhythm with no visible P waves that are replaced by coarse fibrillation waves and irregularly irregular QRS complex.}
	\label{Fig2:AFIB12LeadECG}
\end{figure}

An electrocardiogram (ECG) is a time-series measurement of the electrical signal generated by the human heart from the surface of the human body. This periodic signal consists of three main sections, as shown in Figure~\ref{fig1:ECG waves}: the P wave, QRS complex, and T wave. ECG recordings are an important and widely used tool for heart disease diagnoses. The amplitude, duration and morphological features of the different sections of the ECG signal are used in the diagnoses and classification of dozens of distinct heart conditions. Arrhythmia is a group of conditions in which the heartbeat has an irregular rate or rhythm. For instance, the most common type of arrhythmia is atrial fibrillation, which is characterized by rapid and irregular heartbeats, as depicted in Figure~\ref{Fig2:AFIB12LeadECG}.

Applications of machine learning techniques for automated analysis of ECG data have been the focus of many recent cardiac research efforts such as arrhythmia classification~\cite{ZhengJ2020} and accurate prediction of ventricular arrhythmia origins ~\cite{Zheng2021,Zhengjw2020}. In addition to being the most important tool in cardiology diagnoses, ECGs have the potential to be used as a biometric in human identification systems, similar to fingerprint, face and iris. This concept can be implemented easily and inexpensively ~\cite{Biel,Carreiras} and eliminates the aliveness test required in some other forms of biometrics since a heart signal is an inherently alive biometric. An ECG recording can also be used to estimate the demographic characteristics of the patient. For instance, high accuracies have been reported for age estimation, sex detection and race classification based on ECG signals ~\cite{Attia2019AgeAS,Noseworthy,Cabra2018WideML,Wiggins2008EvolvingAB}. Furthermore, ECG signals have been successfully used for emotion recognition ~\cite{Koelstra,BRAS2018}. 

Although the application of ECG in biometrics, demographic prediction and disease detection seems promising and can have many useful applications, this also raises serious privacy concerns. Additionally, emerging medical wearable technologies make the privacy requirements in the analysis of such data even more critical. Medical wearable devices capture a continuous stream of signals and measurements from our bodies, which will open vast opportunities in the research, monitoring, prevention and diagnosis of diseases in the near future. For instance, Apple recently released Apple Watch series 6 capable of capturing a single lead ECG. Both the sensitivity of ECG data in terms of the personal information that it can reveal about its owner and the fact that we are amassing large amounts of ECG data coming from the wearable sensors, make privacy safeguards in the collection and analysis of such data critical. For example, an ECG biometrics system can also diagnose and store heart conditions. The authors also consider a  possible privacy disclosure scenario not discussed in the ECG data privacy literature: consider an individual who contributes their ECG data to two different research datasets A and B. For instance, database A has an ECG sample, sex and date of birth, while database B has an ECG sample and zip code. By simply matching the ECG columns in both databases using an ECG biometric identification system, one can discover the individuals who appear in both datasets and obtain a complete profile of these subjects by joining their records. In this reidentification scenario, we have sex and date of birth from database A and zip code from database B. These three demographic attributes are enough to uniquely identify 87\% of US citizens ~\cite{Sweeney2,Hayes}. Multiple incidents have occurred in the past two decades, where individuals have been reidentified from anonymized datasets. For instance, a researcher from MIT reidentified the Massachusetts Governor's medical record from an anonymized insurance dataset by joining it with voter registration list auxiliary data ~\cite{Sweeney}.
In contrast to the common traditional belief that statistical aggregate databases are safe to share, it can be proven that aggregation and basic anonymization does not guarantee privacy and that individuals can be reidentified even from the published aggregated results ~\cite{Liuetal2018}. Thus, being able to share aggregated statistics from private datasets while preserving an individual’s privacy is extremely important. In addition to reidentification attacks that trace whether a specific individual’s data are included in a dataset, an adversary  might be able to partially reconstruct a private dataset from public aggregate information, known as a \emph{reconstruction attack}. If the publicly shared statistics are not sufficiently distorted, then an adversary might be able to determine sensitive information about all individuals in the dataset via a reconstruction attack. For example, in a cardiology dataset, the attacker can determine the value for the diabetes attribute (indicating whether a patient has diabetes) for all individuals in the dataset. These reconstructed records can then be reidentified and linked to a specific individual by joining them with other datasets.
In the first part of the paper, we present background information and a survey on differential privacy, focusing on its application to ECG analysis. The second part outlines a detailed approach for performing differential privacy query releases. This includes key activities and decision points involved in publishing differentially private analyses, along with a case study. The guidelines and lessons derived from this study are broadly applicable across various domains and datasets.

\section*{Part I: Literature Review}
In this part, we provide a general introduction to differential privacy, discuss its application to ECG signals, and review related work.
\subsection*{Differential Privacy Background}

The randomized algorithm $M$ is said to be $\epsilon$ differentially-private if for all neighbouring datasets $D$ and $D'$ that differ only on a single element (i.e. the record of one person), and all subsets $S \subseteq R$ of outputs of  $M$,

\begin{center}
$ Pr[M(D) \in S] \leq e^{\epsilon} Pr[M(D') \in S]$
\end{center}

This ensures that addition/removal of a single person to the database does not change the results of the analysis too much. More specifically, the probability that the outcome of an analysis changes after adding an individual's record to the database is guaranteed to be limited by a factor of $e^\epsilon$. This makes the exposed differentially private query engine immune to attacks in the presence of unforeseen auxiliary information ~\cite{Dwork}. There is also an $(\epsilon,\delta)$ differential privacy definition, in which a small chance of accidental information leakage is allowed via the $\delta$ parameter:
\begin{center}
$ Pr[M(D) \in S] \leq e^{\epsilon} Pr[M(D') \in S]+\delta$
\end{center}

In differential privacy, we guarantee the privacy criteria by adding random noise to the original output of a query. Laplace is one of the main mechanisms for adding noise in differential privacy. In the Laplace mechanism, the noise is drawn randomly from a Laplace distribution with mean zero and scale of $\frac{\Delta f}{\epsilon}$. To recall, we define a zero-mean Laplace distribution with a scale of $b$ as
\begin{center}
$Lap(x|b) = \frac{1}{2b} e^{- \frac{|x|}{b}}$
\end{center}

In addition to epsilon, we need to specify the  sensitivity of our function ($\Delta f$): the maximum change in the result of a function $f$ when we add or remove one record to/from the database. We define the $l_1$ sensitivity of a function $f:N^{|\mathscr{X}|}\rightarrow R^{k}$ as
\begin{center}
$$ \Delta f= \max |f(x)-f(y)|\ \ \  x,y \in  N^{\mathscr{|X|}},\ \ \ |x-y| = 1 $$
\end{center}

Note that this sensitivity is the global sensitivity, i.e. the maximum possible change of function $f$ due to addition or removal of one record from the set of all possible records for that database. Global sensitivity is used to guarantee the amount of privacy loss in the worst-case scenario. This is versus the local sensitivity, which is the maximum change in a given function due to addition or removal of one record to/from a specific instance database.

Differential privacy can be applied in two modes: interactive and noninteractive. The interactive mode works by perturbing the results of a query before returning to the analyst on the fly. On the other hand, in the noninteractive setting, also known as query release, the data custodian either publishes some statistics and results on the database or anonymizes and releases a synthetic database. In the case of a synthetic database, users can query the released database as many time as they want, still holding the privacy guarantee. Unfortunately, generating synthetic differentially private datasets can be computationally very complex and a difficult task, especially for high-dimensional data. If the queries are known in advance, then the noninteractive query model can give the best possible utility since it is able to correlate noise~\cite{Dwork}.

It can be proven that there is no private mechanism that can answer an arbitrary number of queries accurately. Dinur\&Nissim~\cite{DinurNissim} prove that in a database of records consisting of bits, if a user wants to know the sum of a random subset of the bits, then no private mechanism can answer n queries with error O($\sqrt{n}$). Thus, a total budget for $\epsilon$ should be allocated for a differentially private database, and once the users exhaust this $\epsilon$ budget, the database needs to be shut down to prevent potential privacy leaks. For example, if a researcher (trusted party) wants to publish the results from a private research database safely to the public (untrusted parties) using differential privacy techniques, then a total $\epsilon$ budget needs to be determined to be used for all queries. This total budget needs to be distributed across different queries used to generate the results. 

Its robustness to postprocessing, composability and graceful degradation in the presence of correlated data makes differential privacy even more attractive since we can build complicated differentially private mechanisms in a modular fashion with a guarantee on the amount of privacy loss in the worst-case scenario. A very useful characteristic of differential privacy is its composability: Given a set of differentially private computations, the overall privacy loss parameter is equal to the sum of epsilons for each query. In other words, if $F_1(x)$ is $\epsilon_1$-differentially private and $F_2(x)$ is $\epsilon_2$-differentially private, then the mechanism $G(x) = (F_1(x), F_2(x))$ that releases both query results on the same input dataset is $\epsilon_1+\epsilon_2$-differentially private.

 The differential privacy concept has also been applied to machine learning techniques to enable the training and sharing of models with privacy guarantees. For instance, one popular method to incorporate differential privacy in the learning process is the gradient perturbation, where we add noise to the results of the gradient of the loss function at each iteration of the algorithm. However, in their current state, most of them require very large values for the privacy budget in order to provide an acceptable utility ~\cite{JayaramanEvans}. 

In the context of differential privacy, storing raw data on a central server and adding noise to the query results before sending back responses is known as global (central) differential privacy. This method ensures that individual data points are protected by introducing uncertainty into the results, but it relies on the assumption of a trusted central authority. However, there are alternative strategies for applying differential privacy that may be more suitable for specific scenarios. One such strategy is local differential privacy (LDP). In LDP, each user's data is perturbed locally on their device before it is sent to the central server. LDP ensures that even if an adversary gains access to individual responses, they cannot learn significant information about any user. This model employs a distributed architecture and assumes an untrusted curator, meaning that the data collection entity may not be fully trustworthy. By perturbing individual responses before transmission, LDP allows for valid statistical inferences while protecting individual privacy.

Another approach within the realm of differential privacy is the generation of differentially private synthetic data. This method involves taking an original dataset \( X \) as input and producing a synthetic dataset \( Y \) that meets differential privacy criteria. The records in this synthetic dataset do not correspond to real individuals, thereby mitigating privacy concerns. The synthetic dataset can be publicly shared, allowing for unlimited queries as with any open dataset. The primary challenge with differentially private synthetic data generation is maintaining accuracy; producing a synthetic dataset that accurately reflects the original data while preserving privacy is a complex task.

\subsection*{Differential Privacy Implementations}

In recent years, in parallel with the theoretical developments in differential privacy, tools have been developed for real-life deployment of differentially private systems. For example, Google developed an open source library to generate $\epsilon$- and ($\epsilon$, $\delta$)-differentially private statistics over datasets with interfaces for Java, Go and C++ programming languages (https://github.com/google/differential-privacy). It assumes that each user contributes only a fixed number of rows to each partition and that number is configurable by the user (the number of times a user appears in a database or maximum number of contributions each user can make to a single aggregation). The core library supports differentially private operations such as sum, mean, count, variance, standard deviation, min, max and median. Additionally, it supports implementations of the general Laplace and Gaussian mechanisms, which can be used to make any given function differentially private. It also includes an SQL engine on top of it to easily perform differentially private analysis on top of tabular data. Privacy on Beam is an end-to-end differential privacy framework in Google's differential privacy library, which is built on top of Apache Beam, making it possible to run differentially private queries at scale and in parallel on large data. Another useful tool included in the Google Differential Privacy project is DP Stochastic Tester, which checks whether a given algorithm holds the differential privacy predicate over a set of databases.

Diffpriv ~\cite{pmlr-v70-rubinstein17a} (https://github.com/brubinstein/diffpriv) is an R library for differential privacy, supporting implementations of generic differential privacy mechanisms, including Laplace, Gaussian, exponential and Bernstein ~\cite{Bernstein}. 
SmartNoise/OpenDP is another open-source differential privacy library  jointly developed by Microsoft and Harvard. The SmartNoise system includes differentially private algorithms, as well as support for connecting to popular SQL engines. It has tools to produce differentially private synthetic datasets generated from a statistical model based on the original dataset. IBM has a differential privacy Python library called Diffprivlib (https://github.com/IBM/differential-privacy-library). In addition to basic differentially private functions such as histogram and mean, it comes with a set of differentially private models including linear regression, logistic regression, k-means, naive Bayes and PCA. In this paper, we use IBM's library to implement the differentially private release of some queries on an ECG dataset.

\subsection*{Differential Privacy in ECG Analysis}
Depending on the usage scenario and setting, we can choose from various types of differential privacy, such as global, local, or hybrid, based on their suitability for the ECG analysis task at hand:
\begin{itemize}
\item \textbf{Global Differential Privacy} is appropriate for hospital settings where the original ECG recordings must be stored in a database. Researchers can access this database via a differentially private query engine, which adds noise to the query results to protect individual privacy.

\item \textbf{Local Differential Privacy} is suitable for medical wearable devices or remote monitoring systems where a trusted central aggregator is not available. Each ECG record is independently perturbed with noise before being sent to a central server for aggregation. This ensures privacy at the source, as the data is protected before leaving the device.
\end{itemize}

Similar to the local differential privacy, in the global differential privacy the perturbation can also be applied to all the records individually in the central server, even though this is less common. In the global Differential privacy setting, it is more practical and effective to follow the standard approach of adding noise at the aggregate query level or using a synthetic data generation approach which offer a better data utility.

In a global differential privacy setting, we can leverage synthetic data generation techniques. Algorithms such as Generative Adversarial Networks (GANs) can create synthetic ECG data that mimics the real data. Differential privacy is enforced during the training process to ensure that the synthetic data does not reveal information about any individual in the original dataset. These synthetic datasets can be shared with external parties, researchers, or the public without risking exposure of sensitive information and can also be used to train machine learning models, protecting them from membership inference attacks. Although analysts can theoretically use synthetic data for any analysis, in practice, synthetic data must be constructed to support specific, predefined target analyses. 

A hybrid approach, incorporating both global and local differential privacy, can be used for training differentially private models, such as in differentially-private federated learning. In this approach, multiple institutions can collaboratively train a machine learning model without sharing their sensitive raw data. For example, multiple hospitals can train a model together by each training a local model on their ECG data and sharing only model updates with a central server. Noise is added to these updates to ensure differential privacy. This method can also be beneficial for medical wearable devices, where model updates from each device are sent to a central server to be aggregated, with the final aggregated updates sent back to each device.

In addition to selecting the setting of your differential privacy such as global vs local, you also have to decide at what point do you want to add noise to your data or results. In ECG analysis, noise can be introduced at various stages to ensure data privacy. These stages include:
\begin{itemize}
\item \textbf{The Original ECG Signal:} Noise can be added directly to the ECG signal, which is treated as time series data. This ensures that the raw signal itself is protected before any further processing or analysis. The perturbed raw signals can be shared with the user to perform their analysis freely. This is a form of non-interactive differential privacy in which a differentially private dataset or summary is released once (single release), and users can run their analyses on this release without further interaction with the original data. In the local differential privacy setting, for instance this can be applied to the ECG signals right after it is captured on the medical wearable device and before sending to a central server.  

\item \textbf{On Extracted Attributes:} Noise can be added to attributes extracted from the raw signals, such as the length of the QRS complex, the heart rate, or other clinically relevant features. This approach protects specific characteristics derived from the ECG data. Similar to adding noise to the original ECG signal, you can do this on a local differential privacy setting, by adding noise to the extracted attributes before sending it to the central server.

\item \textbf{On Aggregate Function Results:} Differential privacy noise can be applied to the results of aggregate functions, such as the mean or sum of certain ECG features across multiple patients. This method is useful for preserving the privacy of individuals when publishing summary statistics. This is usually done in a global differential privacy setting, where we have the original records from all the individuals in a central database.

\item \textbf{During the Training Process of Machine Learning Models:} Differential privacy can be incorporated into the training process of machine learning models. For instance, Differentially Private Stochastic Gradient Descent (DP-SGD) adds noise to the gradient updates during model training, ensuring that the model does not memorize or reveal sensitive information from the training data.
\end{itemize}

Below is a summary of previous work done in privacy-preserving ECG analysis including the application of differential privacy to ECG data. Ma et al.~\cite{Maetal2023} propose a Renyi-differentially private-GAN (RDP-GAN), which achieves differential privacy by adding random Gaussian noise to the value of the loss function during training. The authors claim that it achieved a better privacy level while producing high-quality samples compared with a benchmark DP-GAN (Differentially Private Generative Adversarial Network) which works using perturbation of gradients. Huang et al.~\cite{Huangetal2019} present an ECG-based authentication scheme for IoT-based healthcare, designed to handle noisy ECG signals, such as ECG samples captured when the user is running. They remove noise using the Singular Value Decomposition (SVD) method. Their scheme uses amplitude and time fiducial components of the PQRST wave for authentication matching and protects the privacy of stored ECG templates using the differential privacy Laplace mechanism. Habiba et al.~\cite{Habibaetal2021} introduce a technique to generate synthetic ECG signals using a Generative Adversarial Network (GAN) with a Neural Ordinary Differential Equation (Neural ODE) based Generator and Discriminator. Their approach does not use differential privacy during model training but relies on the protection offered by the synthetic data itself. Jafarlou et al.~\cite{Jafarlouetal2022} propose a GAN-based framework for de-identifying ECG signals, leveraging a combination of standard GAN loss, ODE-based loss, and identity-based loss values. Their generator de-identifies ECG signals while preserving important information regarding cardiovascular conditions, without relying on differential privacy. Ying et al.~\cite{YING2023101568} present a privacy-preserving federated semi-supervised learning framework for predicting ECG abnormalities. They transform ECG signals into 2D images using Gramian Angular Summation/Difference Fields transformation~\cite{Wangetal2015} and employ a ResNet-based architecture. Pseudo-labeling annotates unlabeled data on client devices using the existing global model. Initially, a medical server sends trained global model parameters to clients (e.g., smartwatches), which then use pseudo-labeling on the new local data to label them and then train new local models. These local models are sent back to the central server for aggregation. Liu et al.~\cite{Liuetal2020} propose a novel blockchain-enabled online learning model under local differential privacy for coronary heart disease diagnosis in mobile edge computing. A local differential privacy scheme is used to protect the privacy of patients. Agrawal et al.~\cite{agrawal2024federated} combine federated learning and differential privacy techniques to train an ECG classification model across seven hospitals. Local models are trained using the DP-SGD algorithm to preserve patient privacy at each hospital before being sent to central servers for aggregation. Son et al.~\cite{Sonetal} proposed a patient-worn ECG sensor-based heart monitoring system that also preserves the privacy of users by adopting signal scrambling and anonymous identity schemes. Bonomi et al.~\cite{Bonomietal2022} propose a privacy-protecting method for sharing individual-level ECG time-series data, using dimensional reduction and random sampling to achieve provable privacy protection. Their sanitization method, based on metric privacy~\cite{Chatzikokolakis2013BroadeningTS}, a generalization of differential privacy, uses the Euclidean distance between the first coefficients in the Discrete Cosine Transform (DCT) domain to capture the distance between two ECG time-series. This approach provides strong privacy protection against an informed adversary while enabling useful aggregate-level analysis.

\section*{Part II: Case Study}
In this section, we first outline a step-by-step approach for performing differentially-private query releases. Next, we apply this method to a real-world ECG dataset. We employ a non-interactive global differential privacy setting, utilizing the Laplace mechanism to perturb the results of the summary analysis.

\subsection*{Publishing a Differentially Private Analysis: A Six-Step Approach}

\begin{figure}
    \centering
	\includegraphics[scale =0.6]{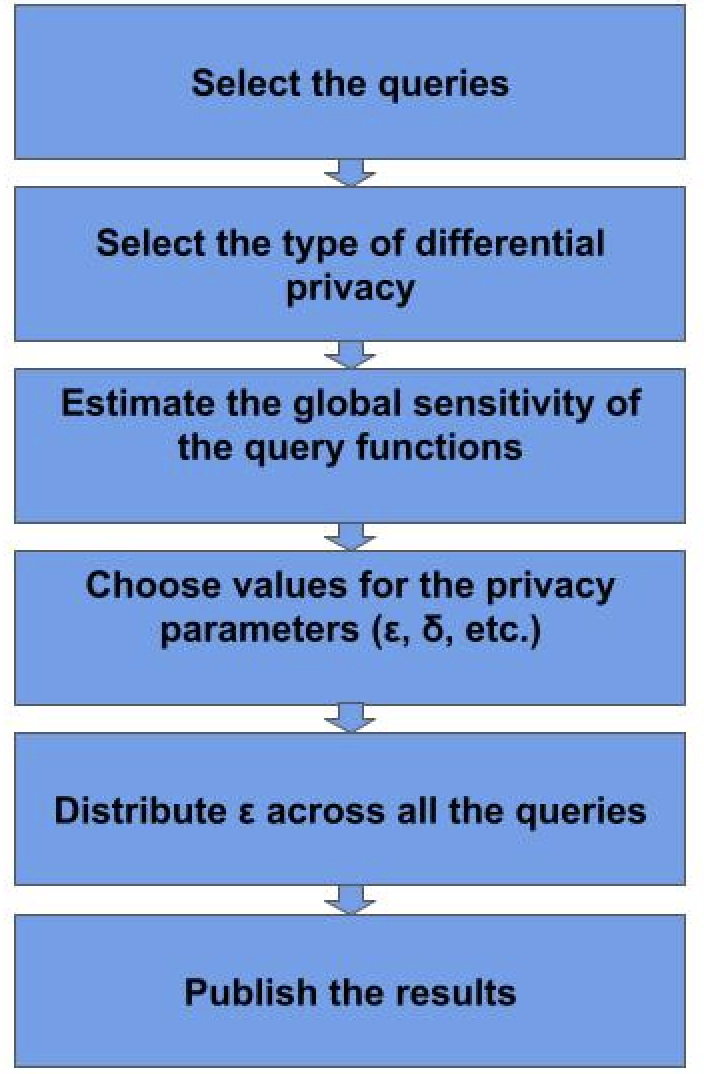}
	\caption{Steps in a noninteractive differential privacy query release}
	\label{queryrelease-steps}
\end{figure}

Query release is the problem of releasing accurate answers to a set of statistical queries in a privacy preserving setting. Many query release algorithms work by generating synthetic data: an approximate/perturbed version of the original database that works on every statistical query of interest. Often, these methods are computationally expensive and intractable. Another method is to just publish the results of some queries and not the database itself. 
 In the coming subsections, we will introduce our six-step approach for releasing differentially private query results, as shown in Figure~\ref{queryrelease-steps}. In the Methods section, we will explain how we applied this same process to the ECG dataset used in our case study.\\

 \textbf{1. Select the Queries: }First, we have to decide on the statistics or queries that we want to share with the public. Considering the limit on the privacy budget ($\epsilon$), we need to find the most important or useful analysis to share. 

\textbf{2. Select the Type of Differential Privacy: }In the Differential Privacy Background section, we provided the definition for two versions of differential privacy depending on whether the $\delta$ parameter is set to zero (pure differential privacy) or not (approximate differential privacy). However, many other relaxed variations of differential privacy have been offered by researchers ~\cite{BalazsDesfontaines}. For instance, ($\epsilon$, $\delta$,$\gamma$)-random differential privacy relaxes the original differential privacy constraint such that adding a \emph{randomly} drawn new observation to a database will have a small effect on output ~\cite{Hall_Wasserman_Rinaldo_2013} (the original defenition of differential privacy requires that adding \emph{any} new observation to a database will have small effect on the output of the data-release procedure). Another example is Kullback-Leibler privacy, in which they use an average-case risk model (the arithmetic mean of the privacy loss random variable) instead of the original worst-case privacy loss version ~\cite{barber2014privacy,CuffYu}. An $(\alpha,\bar{\epsilon})$-Rényi Differential Privacy (RDP) requires that the Rényi divergence of order $\alpha$ between $F(x)$ and $F(x')$ to be bounded by $\bar{\epsilon}$~\cite{Ilyaetal2017}. Rényi divergence of order $\alpha$ is generalization of Kullback-Leibler divergence and is a measure of information that compares two probability measures defined on the same measurable space.
\textbf{3. Estimate the Global Sensitivity of the Query Functions: }
We need to have the global sensitivities early in our process because it is needed in the next steps such as choosing values for the epsilon and other parameters. Estimating a global sensitivity for a given general function $f$  can be challenging. Relaxed and easier-to-estimate versions of the sensitivity have been proposed by different researchers. Nissim et al.~\cite{NissimRaskhodnikova} introduced the smooth sensitivity technique which works by estimating an upper bound on the local sensitivity using a smooth function. This helps to reduce the amount of noise added, since the local sensitivity is often much lower than the global sensitivity. Local sensitivity is defined as the maximum change in the result of applying a given function in the neighbourhood of an instance database.

For basic statistics such as mean queries, most differential privacy libraries such as Google differential privacy, Diffprivlib, and Smartnoise require the range/bounds of values for variables to be specified, which is used to estimate the sensitivity by bounding it. The sensitivity for the mean function is $\Delta f=\frac{b-a}{n}$, where  $a$ is the lower bound and $b$ is the upper bound for the values. 
To estimate global sensitivity in differential privacy, one common technique is to transform queries with unbounded sensitivity into equivalent queries with bounded sensitivity through a process called clipping. Clipping involves setting upper and lower bounds on attribute values to restrict the range. For instance, ages above 125 can be clipped to 125, and ages below zero can be clipped to zero. This ensures that all age values fall within the range of 0 to 125. The primary challenge in clipping is determining appropriate upper and lower bounds. For ages, this is straightforward: no one can have an age less than 0, and it is highly unlikely for someone to be older than 125. However, in other domains, identifying these bounds can be more complex. Once clipping is performed, the sensitivity of a summation query on the clipped data is equal to the difference between the upper and lower bounds used. As a general guideline, clipping bounds should ideally encompass 100\% of the dataset or as close to it as possible. If the clipping bounds are determined by examining the data, they might inadvertently reveal information about the dataset. Therefore, clipping bounds are typically set using properties of the dataset that can be known a priori (such as the plausible range of ages) or by performing differentially private queries to iteratively evaluate different clipping bounds. In the latter approach, the lower bound is typically set to 0, and the upper bound is gradually increased until the query's output stabilizes, indicating that no new data points fall outside the current bounds. This method ensures that the bounds are set appropriately without compromising privacy~\cite{near_abuah_2021}.

IBM's diffprivlib requires you to set the sensitivity manually when the general differential privacy mechanisms such as Laplace are being used directly. The library uses a sensitivity sampling technique in place of theoretical sensitivity analysis to achieve ($\epsilon$, $\delta$,$\gamma$)-random differential privacy~\cite{Hall_Wasserman_Rinaldo_2013}. The sensitivity sampler requires that a sampling distribution be provided.
 
With regard to differential privacy, distributions of variables in the dataset can make them good or bad candidates for different types of queries by affecting the sensitivity. Skewed data might make the sensitivity very high for some types of queries, such as the mean; therefore, the amount of noise added can be very large, thus reducing the accuracy of the response. Although the case for normal distributions is better than the skewed data, it can still noticeably distort the results of the mean function. In general, the long tails in the distribution of variables can significantly reduce the utility of analysis. For the mean function, the sensitivity also depends on the size of the database: The larger the number of participants in the database, the smaller the amount of required additive noise. However, the case for count and histogram queries is better since the global sensitivity is always one and independent of the dataset. Especially when we have a large number of users, count or histogram queries can be published with a high utility while still ensuring the worst-case scenario privacy loss amount.

\textbf{4. Choosing Values for the Privacy Parameters: }Despite the vast literature on differential privacy, less attention has been paid to the important question of choosing a proper value for $\epsilon$. Dwork\&Roth ~\cite{Dwork} mentioned that choosing $\epsilon$ is a social question. Lee\&Clifton~\cite{Lee2011HowMI} used a Bayesian approach to find an upper bound for the value of $\epsilon$ based on how much the posterior belief regarding an individual's participation in a database is updated. They have explored the problem of choosing an optimal $\epsilon$ from an adversarial model point of view. Naldi\&D’Acquisto~\cite{Naldi2015DifferentialPA} propose using the confidence level and interval instead of $\epsilon$ because they are more intuitive and easier to understand. These two parameters tell us the probability (confidence level) that the result is within a given range (confidence interval). In other words, $\epsilon$ is chosen to meet the desired level for accuracy. The authors have provided the formulas to convert these two parameters to $\epsilon$ for counting queries. Kohli\&Laskowski~\cite{Kohli} promoted the incorporation of societal preferences in choosing $\epsilon$ based on the privacy preferences of the data contributors. In their approach (called epsilon voting), each data contributor expresses their desired value for $\epsilon$, and a chooser mechanism aggregates all of the users’ preferences into a single final value.

In addition to academic environments, differential privacy has also gained attention in industry. Companies such as Google, Apple and Microsoft are using differentially private mechanisms for collecting and analysing their users' data. For instance, Apple's iOS apps add random noise to personal data such as emoji usage or HealthKit data before storing it for aggregate analysis purposes. By reverse engineering the iOS apps, researchers were able to determine how these apps implement differential privacy and what specific values of $\epsilon$ have been used. Based on their research, values such as 6, 14 and even 43 were used, which are considered to be unsafe (values below one are usually considered as safe) ~\cite{Greenberg}. Google has also started using differentially private algorithms for its data collection and analysis; for instance, chrome leveraged a differential privacy system called Randomized Aggregatable Privacy-Preserving Ordinal Response (RAPPOR) for some time. They used an $\epsilon$ value of two on average and an upper limit of 8 or 9 over the lifetime of the user ~\cite{Erlingsson}. Both Apple and Google have used local differential privacy in their products.
Table~\ref{epsilonvalues} summarizes some of the values used for $\epsilon$ in different academic papers or software. In most cases, the values for $\epsilon$ have been chosen without any justification or reasoning. To determine the proper value for $\epsilon$, we also need to know the level of accuracy needed in that study, for example, what is the minimum accuracy needed to make a report useful for cardiologists. 

\begin{longtable}{|p{5cm}|p{2cm}|p{5.5cm}|}
\caption{Example epsilon values used in research and industry}
\label{epsilonvalues}\\
\hline
\textbf{Product/Research}  & 
\textbf{Epsilon}   &
\textbf{Application} 
\endhead \hline

Apple iOS  & 6, 7, 14 & HealthKit, emoji usage, browsing   \\ \hline
Google Android RAPPOR &  4  &Client-sides statistics collection, chrome usage statistics    \\ \hline
Microsoft Windows 10 ~\cite{Dingetal} & 0.1-1  &   Application usage statistics  \\ \hline
Korolova et al.~\cite{Korolova2009ReleasingSQ}&ln2, ln5, ln10 &Click counts\\ \hline
Machanvajjhala et al.~\cite{Machanavajjhalaetal} & 0.5-3&Recommender system\\
\hline

Cormode et al.~\cite{Cormodeetal}&0.1-1&Location data\\ \hline
Acs\&Castellucia~\cite{Acs-Castellucia}&1&Smart electric metre\\ \hline
Bhasker et al.~\cite{Bhaskaretal}&1.4&Frequent items\\ \hline
Uhler et al.~\cite{Uhler_Slavkovic_Fienberg_2013}&0.1-0.4&Genome data\\ \hline
\end{longtable}

Hsu~\cite{Hsu} et al. proposed an economic method for choosing $\epsilon$  considering the interests of two parties: the data analyst and the prospective participants in a database. These two parties have conflicting interests. On the one hand, the data analyst is concerned about the accuracy of the analysis results; on the other hand, the participants are concerned only about the risk of participation (harm due to potential privacy leakage) vs. the monetary benefits. The researchers considered a scenario where participants are being monetarily compensated for contributing their data, and participants are rational, i.e. they will agree to contribute their private data if the expected benefits outweigh the risks of bad events (privacy leakage). The authors argue that $\epsilon$  is insufficient to model the real-world complexities of conducting a study; instead, they propose using four parameters from which $\epsilon$ can be calculated. These four parameters are $\alpha$ (accuracy level), $A_M(\epsilon,N)$ (measure of accuracy), $B$ (study budget) and $E$ (individual’s expected cost if they \emph{do not} participate in a study). After plugging these parameters into their models, it tells you if an experiment is feasible or not and if yes what is the optimal $\epsilon$ value.

In case one picks the approximate $(\epsilon,\delta)$ differential privacy version in the previous step, then the value of $\delta$ should be less than the inverse of any polynomial in the size of the database. The chances of privacy leak increase with the size of the database; thus, a value less than the inverse of the size of the database should be chosen for $\delta$. The $\delta$ parameter puts an upper bound on the probability at which a differentially private mechanism is allowed to fail. This helps save the privacy budget; however, it breaks the original promise of worst-case scenario protection in differential privacy.

\textbf{5. Distribute $\epsilon$ Across All the Queries: }In a scenario where we have multiple queries, a total budget for $\epsilon$ needs to be allocated and distributed across all the queries selected in the first step. Techniques exist that can improve the accuracy of the analysis once you know the queries ahead of time, allowing more queries per a given  $\epsilon$ budget ~\cite{LiHay,Hardtetal,DworkNaReRoVa09}. Dwork et al.~\cite{DworkNaReRoVa09} presented a mechanism that is $(\epsilon,\delta)$-differentially private with a better utility than the independent Laplace mechanism for the nonadaptive case, where the queries are known ahead of time. Li et al.~\cite{LiHay} and Moritz\&Talwar~\cite{Hardtetal} also proposed methods to reduce noise when all queries are known in advance.

When distributing the privacy budget, it is crucial to consider the presence of minority groups within a dataset. Differential privacy often requires adding a significant amount of noise to protect the privacy of individuals in these minority groups, which can compromise the utility of the analysis results for these groups. For example, this challenge makes it impractical to apply differential privacy to the analysis of rare diseases. If left unaddressed, this issue can lead to health inequities for subpopulations such as racial and ethnic minorities. One approach to mitigate this problem is to assign a slightly higher privacy budget to minority groups. This allocation should be proportional to the sensitivity and size of each subgroup, thereby providing better protection and more accurate results for these groups without overly compromising their privacy. This weighted approach helps balance the trade-off between privacy and utility, ensuring more equitable outcomes across different subpopulations.

\textbf{6. Publish the Results: } Finally, we can run the queries using the chosen parameters in the steps above and share the results with the public. Usually $\epsilon$  value is published as part of the differentially private report.

\section*{Results}

In this section, we will discuss the results of our differentially private query release on an arrhythmia dataset. An open research dataset containing 10-second recordings of 12-lead ECG with a 500 Hz sampling rate from 10,646 patients is used ~\cite{zheng2020}. It features 11 common rhythms and 67 cardiovascular conditions labelled by experts. Table~\ref{dataset-fig} shows sample rows from this database (refer to Table ~\ref{DATASETCOLUMNSTABLE} for the description of columns).

\begin{table}[]
\centering
\caption{Sample records from dataset}
\label{dataset-fig}
\resizebox{\textwidth}{!}{%
\begin{tabular}{lllllllllllllll}
\textbf{Rhythm} &
\textbf{Beat} &
\textbf{\makecell{Patient\\Age }}&
\textbf{Sex} & \textbf{\makecell{Ventricular\\Rate}} & \textbf{\makecell{Atrial\\Rate}} & \textbf{\makecell{QRS\\Duration}} & \textbf{\makecell{QT\\Interval}} & \textbf{\makecell{QT\\Corrected}} & \textbf{\makecell{R\\Axis}} & \textbf{\makecell{T\\Axis}} & \textbf{\makecell{QRS\\Count}} & \textbf{\makecell{Q\\Onset}} & \textbf{\makecell{Q\\Offset}} & \textbf{\makecell{T\\Offset}} \\
AFIB            & RBBB TWC      & 85                  & MALE            & 117                      & 234                 & 114                  & 356                 & 496                  & 81             & -27            & 19                & 208             & 265              & 386              \\
SB              & TWC           & 59                  & FEMALE          & 52                       & 52                  & 92                   & 432                 & 401                  & 76             & 42             & 8                 & 215             & 261              & 431              \\
SA              & NONE          & 20                  & FEMALE          & 67                       & 67                  & 82                   & 382                 & 403                  & 88             & 20             & 11                & 224             & 265              & 415              \\
SB              & NONE          & 66                  & MALE            & 53                       & 53                  & 96                   & 456                 & 427                  & 34             & 3              & 9                 & 219             & 267              & 447              \\
AF              & STDD STTC     & 73                  & FEMALE          & 162                      & 162                 & 114                  & 252                 & 413                  & 68             & -40            & 26                & 228             & 285              & 354              \\
SB              & NONE          & 46                  & FEMALE          & 57                       & 57                  & 70                   & 404                 & 393                  & 38             & 24             & 9                 & 225             & 260              & 427              \\
AFIB            & TWC           & 80                  & FEMALE          & 98                       & 86                  & 74                   & 360                 & 459                  & 69             & 83             & 17                & 215             & 252              & 395              \\
SR              & NONE          & 46                  & MALE            & 63                       & 63                  & 90                   & 376                 & 384                  & 24             & 38             & 11                & 221             & 266              & 409             
\end{tabular}}
\end{table}

We published the following differentially private reports on the ECG dataset:
\begin{itemize}
  \item The mean QRS durations per arrhythmia 
  \item The median QRS durations per arrhythmia
  \item The histogram of variables in the dataset
\end{itemize}
We have provided both mean and median reports, since in some scenarios such as skewed distributions or in the presence of outliers the median is a better statistic and might provide better utility compared to the mean when using differential privacy. In addition to publishing the above reports, we  also trained differentially private machine learning models to classify records to an arrhythmia type.

Table~\ref{queryrelease-qrs-arrhythmia-resultsTABLE} shows the actual and $\epsilon$-differential private mean QRS durations per arrhythmia. Table~\ref{queryrelease-qrs-arrhythmia-median-resultsTABLE} shows the actual and $\epsilon$-differential private median QRS durations per arrhythmia. We have shown the results from two runs to show how they are different across different execution of the queries due to the randomness in the process of adding noise. One observation is that for less frequent conditions such as atrioventricular node reentrant tachycardia in our dataset, the differential privacy library returned 18 or 256. The reason is that since the sensitivity of a mean function has an inverse relationship with the number of subjects $n$, the amount of noise added to the results for the less frequent categories is very large. This has led to results being outside the range for the average QRS duration (the range for QRS duration in our dataset  is 18-256). Most differential privacy implementations clip the numbers that are outside the range to the minimum or maximum value. This indicates that one cannot publish a report with acceptable accuracy on rare conditions.

\begin{figure}%
    \centering
    \subfloat[\centering Actual]{{\includegraphics[scale =0.5]{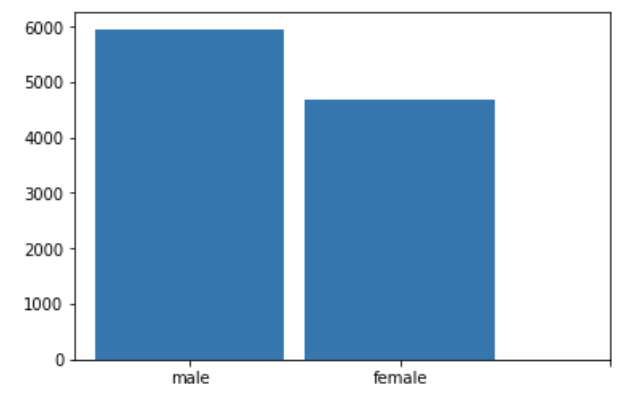} }}%
    \qquad
    \subfloat[\centering Differentially private]{{\includegraphics[scale=0.5]{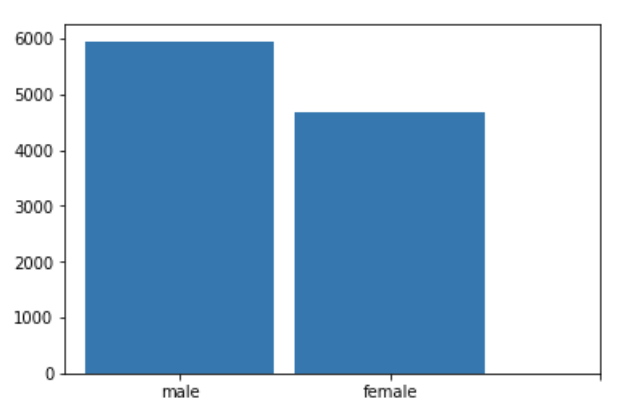} }}%
    \caption{Sex distribution in dataset ($\epsilon=0.2$ for all histograms) }%
    \label{gender-distrib-fig}%
\end{figure}

\begin{figure}%
    \centering
    \subfloat[\centering Actual]{{\includegraphics[scale =0.5]{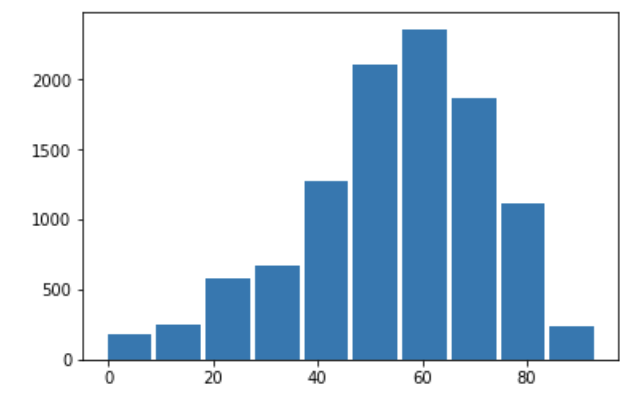} }}%
    \qquad
    \subfloat[\centering Differentially private]{{\includegraphics[scale=0.5]{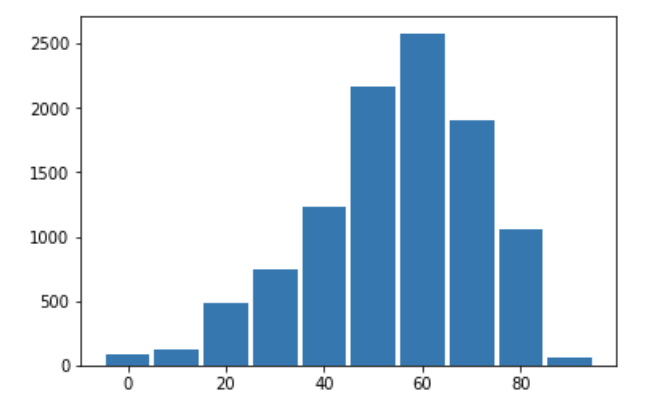} }}%
    \caption{Age distribution in dataset ($\epsilon=0.2$ for all histograms)}%
    \label{age-distrib-fig}%
\end{figure}

\begin{figure}%
    \centering
    \subfloat[\centering Actual]{{\includegraphics[scale=0.48]{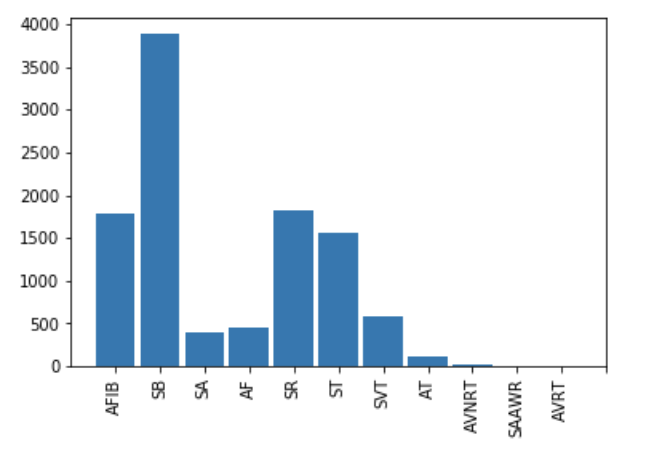} }}%
    \qquad
    \subfloat[\centering Differentially private]	{{\includegraphics[scale=0.48]{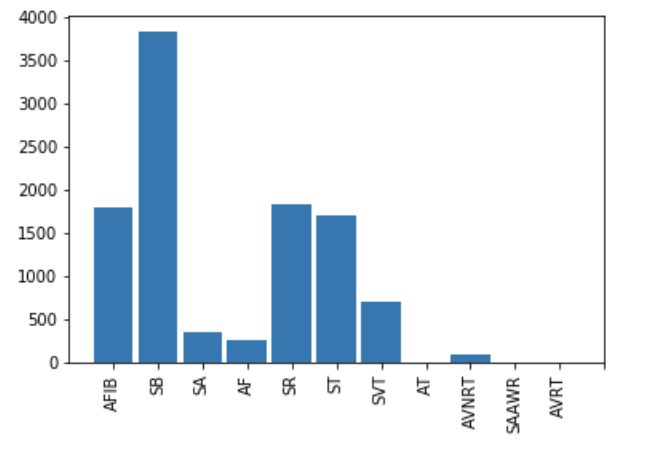} }}%
    \caption{Arrhythmia type distribution in dataset ($\epsilon=0.2$ for all histograms)}%
    \label{arrhythmia-distrib-fig}%
\end{figure}

\begin{figure}
    \centering
	\includegraphics[scale =0.52]{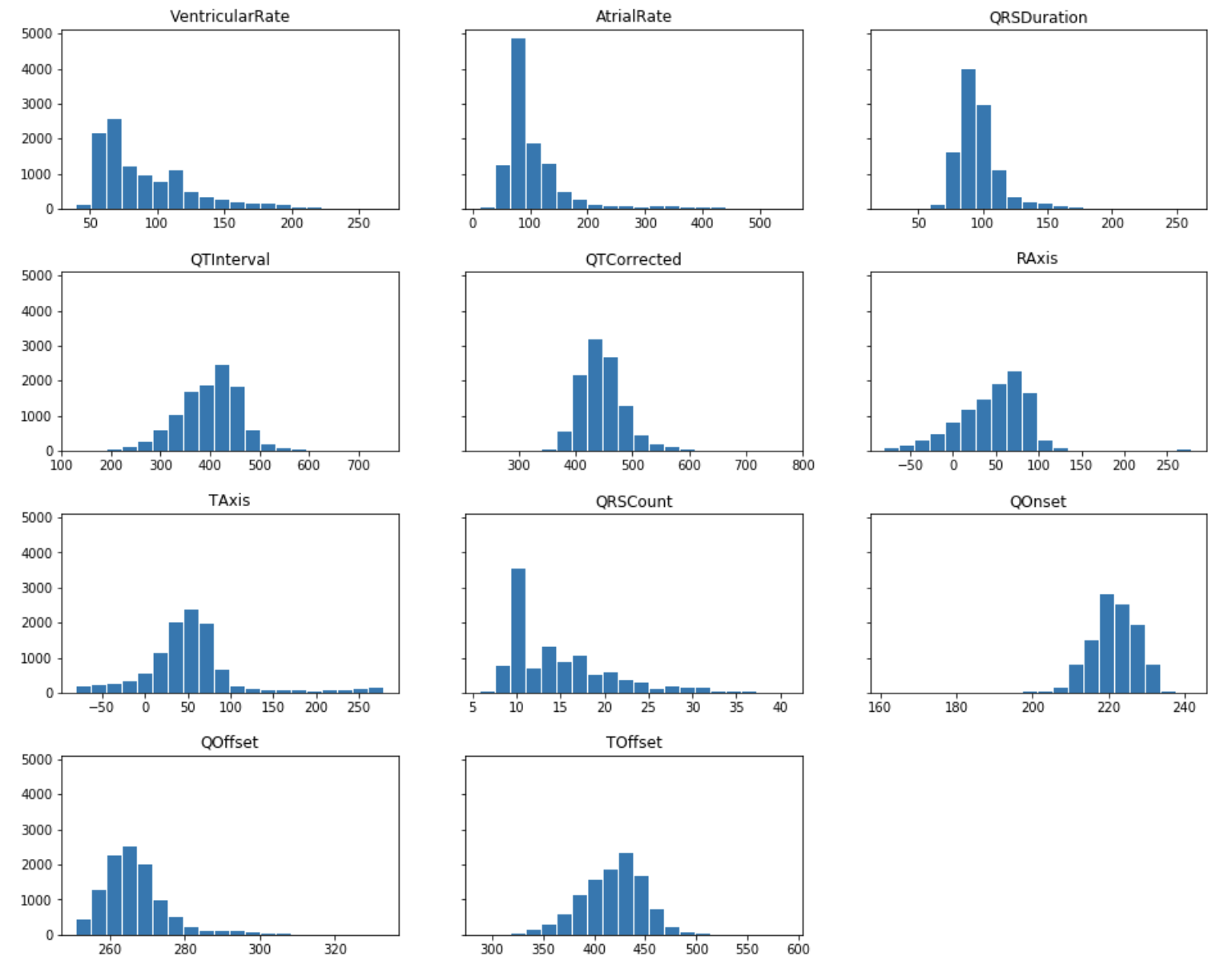}
	\caption{Distribution of ECG variables in dataset}
	\label{ecg-distrib-fig}
\end{figure}

\begin{figure}
    \centering
	\includegraphics[scale =0.52]{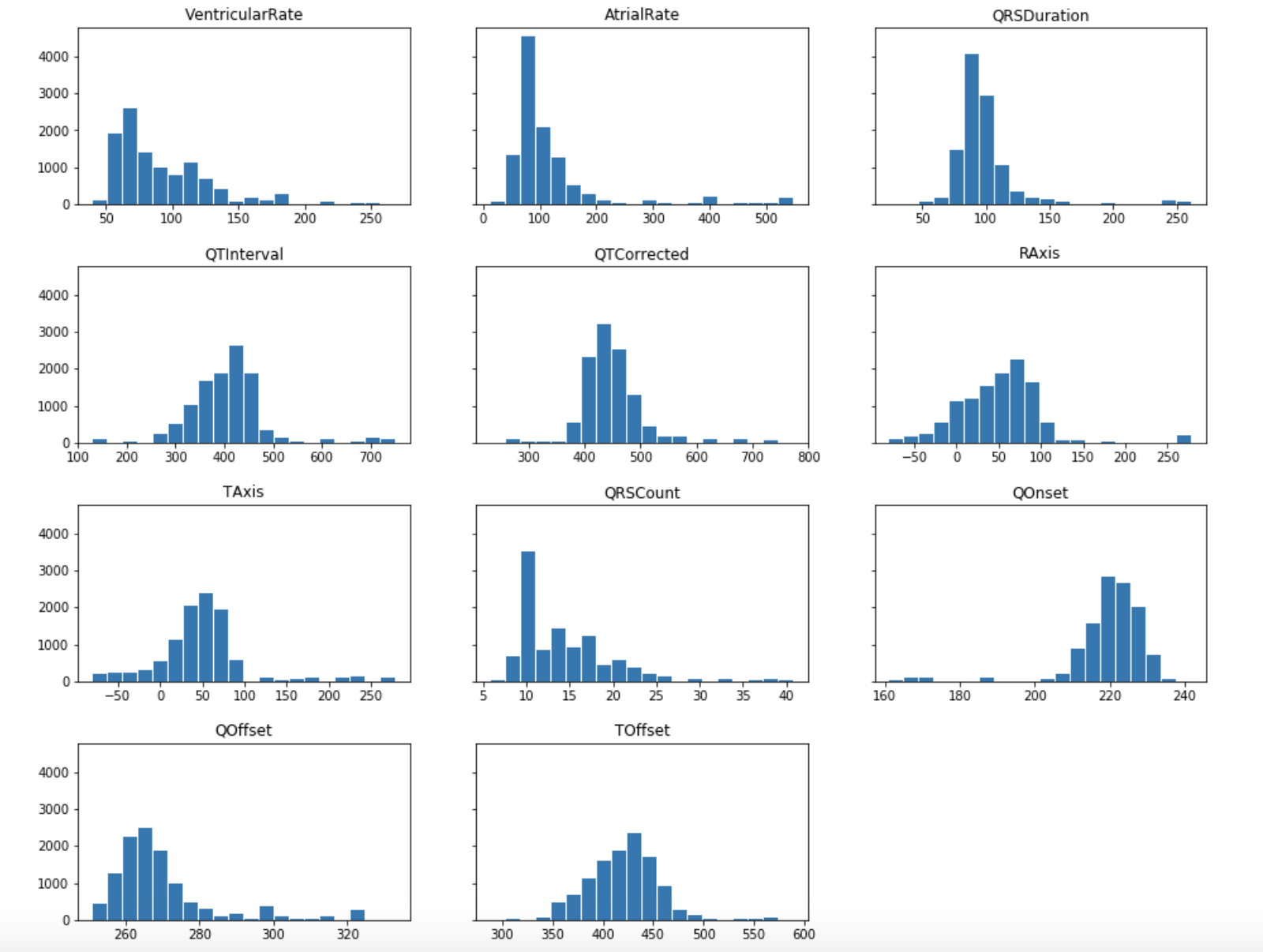}
	\caption{Differentially private distribution of ECG variables in dataset ($\epsilon=0.2$ for all histograms)}
	\label{ecg-dp-distrib-fig}
\end{figure}

Figures~\ref{gender-distrib-fig}, ~\ref{age-distrib-fig} and ~\ref{arrhythmia-distrib-fig} show the actual and differentially private histograms for sex, age and arrhythmia types. Figure~\ref{ecg-distrib-fig} shows the actual histogram for the ECG-related variables, and Figure~\ref{ecg-dp-distrib-fig} shows the differentially private version of it. As you can see, the differentially private histograms are very similar to their corresponding actual histograms.  Note that each of these  reports (the mean, median and histogram reports) consumes 0.2 $\epsilon$ budget. Thus, based on the composability property of differential privacy, the total $\epsilon$ budget for all of these reports is 0.6.
As mentioned earlier, distributions of variables affect the sensitivity, and skewed data might make the sensitivity very large for some types of queries such as the mean. Age and sex are two common fields seen in most cardiology datasets. As seen in  Figure~\ref{gender-distrib-fig} and Figure~\ref{age-distrib-fig} in our dataset, sex has a roughly uniform distribution between males and females, and age resembles a left-skewed distribution (arrhythmia is less common among younger people). 

\newpage

\begin{longtable}{|l|l|}
\caption{Dataset Columns Descriptions}
\label{DATASETCOLUMNSTABLE}
\\
\hline
\textbf{Attribute}  & 
\textbf{Description}
\endhead \hline
Rhythm & Rhythm Label \\ \hline
Beat & Other conditions Label\\ \hline
Patient Age & Age\\ \hline
Sex & Sex (Male/Female) \\ \hline
Ventricular Rate & Ventricular rate in BPM\\ \hline
Atrial Rate  & Atrial rate in BPM\\ \hline
QRS Duration & QRS duration in millisecond\\ \hline
QT Interval & QT interval in millisecond\\ \hline
QT Corrected &QT Corrected QT interval in millisecond\\ \hline
R Axis &R axis\\ \hline
T Axis &T axis\\ \hline
QRS Count &QRS count\\ \hline
Q Onset & Q onset (In samples)\\ \hline
Q Offset & Q offset (In samples)\\ \hline
T Offset & T offset (In samples)\\ \hline
\end{longtable}

Figure~\ref{arrhythmia-distrib-fig} shows the distribution of different types of arrhythmia among the patients (refer to Table~\ref{ARRHYTHMIATYPESTABLE} for a definition of the arrhythmia types). As shown, some form of arrhythmia is rare; for instance, Sinus Atrium to Atrial Wandering Rhythm (SAAWR) is far less frequent than other conditions. In the presence of rare diseases in a dataset, a large amount of noise needs to be added to the results of the analysis to protect the privacy of the participants, significantly decreasing the accuracy of the analysis results. For features related to the PQRST wave, different distribution shapes such as right-skewed, left-skewed, or normal distributions with different Kurtosis values are depicted in Figure~\ref{ecg-distrib-fig}. 

\begin{figure}
    \centering
	\includegraphics[scale =0.8]{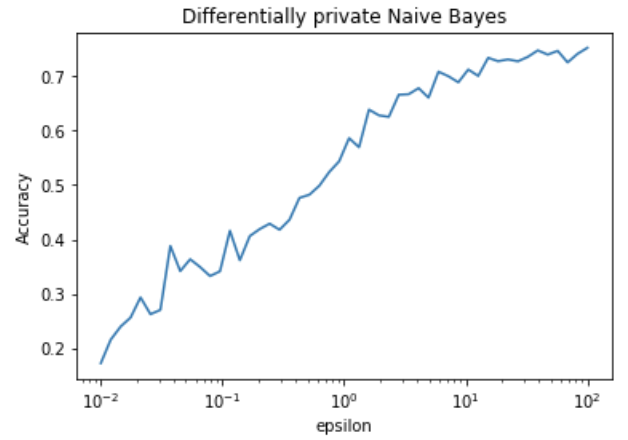}
	\caption{Accuracy of naive Bayes classifier across a range of different $\epsilon$ values}
	\label{epsilon-dp-bayes-classifier}
\end{figure}

\begin{figure}
    \centering
	\includegraphics[scale =0.8]{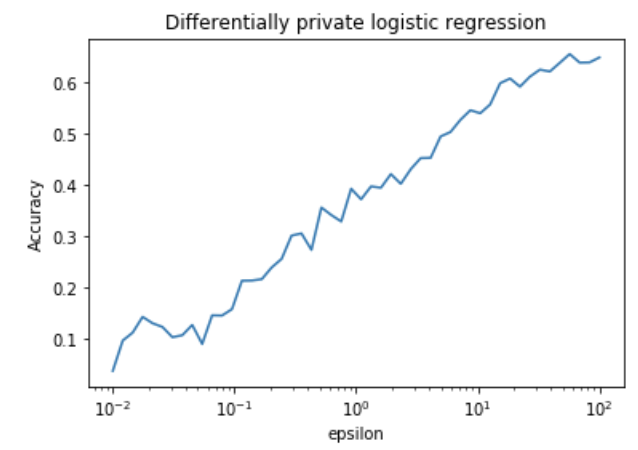}
	\caption{Accuracy of logistic regression classifier across a range of different $\epsilon$ values}
	\label{epsilon-dp-logistic-classifier}
\end{figure}

We also trained machine learning classifiers on our ECG dataset to classify records into eleven groups of arrhythmias specified in Table~\ref{ARRHYTHMIATYPESTABLE}. The goal of this experiment was to explore the performance drop in the differentially private model by comparing its accuracy to the same model but without adding noise. Thus, we chose two common and popular machine learning techniques: naive Bayes and logistic regression. We used the same dataset of 10,646 patients and leveraged 10-fold cross validation to evaluate the model. For naive Bayes, the accuracy achieved using the nondifferentially private version is 74.2\%. Figure~\ref{epsilon-dp-bayes-classifier} shows the accuracies achieved across different $\epsilon$ values. As $\epsilon$ increases, the accuracy approaches that of the regular nondifferentially private classifier. The sharp valleys in the diagram show the nondeterministic behaviour of differentially private algorithms. For the logistic regression model, the accuracy achieved using the nondifferentially private version was 75.80\%. Figure~\ref{epsilon-dp-logistic-classifier} shows the accuracies achieved across different $\epsilon$ values.

As mentioned earlier, without safeguards and precautions Deferentially private analysis or learning algorithms might lead to unfair predictive parity among subgroups. There have been a few efforts in the community to address this issue. Tran et al.~\cite{Tran2023FairDPCF} introduced FairDP, a mechanism designed to achieve certified fairness by independently training models for distinct individual groups using group-specific clipping terms. Jagielski et al.~\cite{Jagielski2018DifferentiallyPF} offered a differentially private learning mechanism which  s guarantee approximate notions of statistical fairness across the groups such as Equalized Odds ~\cite{Hardt2016EqualityOO}.

\newpage

\begin{longtable}{|p{7cm}|p{1.1cm}|p{2cm}|p{2cm}|}
\caption{Actual and $\epsilon$-Differentially private mean QRS durations per arrhythmia ($\epsilon$=0.2)}
\label{queryrelease-qrs-arrhythmia-resultsTABLE}
\\
\hline
\textbf{Arrhythmia Type}  & 
\textbf{Actual}  & 
\textbf{Differentially Private Run 1} &
\textbf{Differentially Private Run 2}
\endhead \hline
Atrial Fibrillation& 92.7809& 100.3062& 81.6582\\ \hline
Sinus Bradycardia& 93.3181& 93.4313& 92.2033\\ \hline
Sinus Irregularity& 87.4536& 18.0000& 145.1477\\ \hline
Atrial Flutter& 97.2989& 127.7923& 176.5641\\ \hline
Sinus Rhythm& 87.0044& 87.8604& 92.4898\\ \hline
Sinus Tachycardia& 85.2768& 76.0064& 87.0250\\ \hline
Supraventricular Tachycardia& 96.0545& 85.7767& 96.3541\\ \hline
Atrial Tachycardia& 88.9587& 52.0237& 18.0000\\ \hline
Atrioventricular Node Reentrant Tachycardia& 89.8750& 18.0000& 256.0000\\ \hline
Sinus Atrium to Atrial Wandering Rhythm& 84.8571& 256.0000& 256.0000\\ \hline
Atrioventricular Reentrant Tachycardia& 81.5000& 120.5715& 256.0000\\ \hline
\end{longtable}

\begin{longtable}{|p{7cm}|p{1.3cm}|p{2cm}|p{2cm}|}
\caption{Actual and $\epsilon$-Differentially private median QRS duration per arrhythmia ($\epsilon$=0.2)}
\label{queryrelease-qrs-arrhythmia-median-resultsTABLE}
\\
\hline
\textbf{Arrhythmia Type}  & 
\textbf{Actual}  & 
\textbf{Differentially Private Run 1} &
\textbf{Differentially Private Run 2}
\endhead \hline
Atrial Fibrillation& 88.0000& 89.8094& 89.3675\\ \hline
Sinus Bradycardia& 92.0000& 91.6741& 88.3865\\ \hline
Sinus Irregularity& 86.0000& 216.8734& 87.1713\\ \hline
Atrial Flutter& 90.0000& 219.0696& 183.2532\\ \hline
Sinus Rhythm& 86.0000& 84.1707& 87.1782\\ \hline
Sinus Tachycardia& 82.0000& 81.1739& 85.6976\\ \hline
Supraventricular Tachycardia& 84.0000& 83.7181& 99.5226\\ \hline
Atrial Tachycardia& 86.0000& 75.9621& 79.5899\\ \hline
Atrioventricular Node Reentrant Tachycardia& 82.0000& 170.2546& 203.1612\\ \hline
Sinus Atrium to Atrial Wandering Rhythm& 90.0000& 44.5616& 80.2591\\ \hline
Atrioventricular Reentrant Tachycardia& 78.0000& 255.0393& 126.6465\\ \hline

\end{longtable}

\section*{Discussion}
while the application of differential privacy to ECG data follows the same fundamental principles as its application to other datasets, the unique aspects of medical data require focused solutions to address correlations, familial dependencies, skewed distributions, rare conditions, and the critical need for accuracy in medical analysis.

The specific characteristics of medical and ECG data introduce several important considerations:
\begin{itemize}
\item \textbf{Correlation Among Fields:} ECG data often includes multiple interdependent variables. The correlations between these variables must be preserved to maintain the utility of the data while still ensuring privacy. This complexity requires advanced techniques to add noise in a manner that respects these correlations. Differential privacy is often critiqued for its vulnerability to data correlation, particularly in medical datasets where genetic and hereditary factors play a significant role. For example, heart conditions can affect multiple family members due to shared genes. In contrast to the earlier belief, recent research has shown that data correlation poses a substantial privacy threat, leading to various types of privacy leakages~\cite{biswas2021machine}. Gehrke et al.~\cite{gehrke2011towards} highlighted these issues in the context of social networks, where user data is highly correlated. They demonstrated that even strong privacy guarantees provided by differential privacy could not fully protect against privacy breaches in such settings. Following this, researchers have continued to explore the privacy risks associated with data correlation and have proposed solutions to mitigate these threats. Another issue related to data correlation is its impact on sensitivity. Correlated data increases the sensitivity of the dataset, which, when applying differential privacy algorithms, results in the addition of more noise to the original data.

\item \textbf{Family Members in the Dataset:} Medical datasets can include data from related individuals, which introduces dependencies that can affect privacy guarantees. Differential privacy mechanisms must account for these dependencies to prevent potential privacy breaches due to familial similarities.
\item \textbf{Skewed Distributions:} Medical data, including ECG records, frequently exhibits skewed distributions, with certain values being much more common than others. Differential privacy algorithms must be tailored to handle these distributions effectively, ensuring that the noise added does not disproportionately affect the most common values.

\item \textbf{Rare Diseases:} In medical datasets, certain conditions or diseases may be rare, resulting in very few records. Protecting the privacy of individuals with rare conditions while allowing meaningful analysis is a significant challenge. Differential privacy needs to ensure that noise addition does not completely obscure the presence of these rare events. Minority groups for which we have fewer subjects in the dataset, such as patients with rare diseases, younger or older patients, and females, might be exposed to health inequity because more noise is required for these groups to protect their privacy. This will make the results reported for these minority groups less accurate and in many cases useless. There have been many reports in recent years on how the implementation of differential privacy in sharing data and statistics will produce false beliefs about health and other crucial aspects of racial and ethnic minorities. Smaller subpopulations require more noise to distort the numbers to guarantee the same level of privacy as the larger populations ~\cite{Santos-Lozada13405,WezerekRiper2020}.

\item \textbf{Sensitivity of the Medical Field:} The accuracy of analyses conducted on medical data is critical, as it can directly impact diagnoses and patient outcomes. Thus, maintaining a high level of data utility while applying differential privacy is paramount. Any noise added to ensure privacy must be carefully calibrated to avoid compromising the effectiveness of clinical decisions.
\end{itemize}

Although it has a strong theoretical foundation, differential privacy has many unaddressed questions regarding how to estimate its parameter values such as $\epsilon$. Additionally, an acceptable privacy guarantee requires a large loss in utility.  In our experience, in its current state, an analyst should distinguish between what is publishable differentially private vs. what they would like to publish in an ideal world. For example, it can be practical for some types of reports such as count or histogram queries with a large number of subjects in a noninteractive mode. Random differential privacy techniques seem to be more practical than the pure $\epsilon$-differential privacy as they add less noise to the results, but apparently they do not guarantee the initial worst-case scenario privacy loss promise of differential privacy. New and better techniques are required for a more widespread application to cardiology and in general medical research. The recent open-source implementations of differential privacy in popular languages such as Python, R and SQL open the door for the community to apply this technique to real-world data and applications. This will provide useful feedback to researchers in terms of the areas to improve to make differential privacy more applicable and useful in practice.

\begin{longtable}{|l|l|}
\caption{Arrhythmia Code Mapping}
\label{ARRHYTHMIATYPESTABLE}
\\
\hline
\textbf{Acronym}  & 
\textbf{Full Name}
\endhead \hline
SB & Sinus Bradycardia \\ \hline
SR & Sinus Rhythm \\ \hline
AFIB & Atrial Fibrillation \\ \hline
ST & Sinus Tachycardia \\ \hline
AF & Atrial Flutter \\ \hline
SA & Sinus Irregularity  \\ \hline
SVT & Supraventricular Tachycardia \\ \hline
AT & Atrial Tachycardia \\ \hline
AVNRT & Atrioventricular Node Reentrant Tachycardia \\ \hline
AVRT & Atrioventricular Reentrant Tachycardia \\ \hline
SAAWR & Sinus Atrium to Atrial Wandering Rhythm \\ \hline
\end{longtable}

Another challenge is the complexity of the notion of $\epsilon$ for the end users of the report. $\epsilon$ is not a familiar concept for most people, making the interpretation of the shared results difficult in terms of reliability and accuracy. Additionally, $\epsilon$ is a new notion for the judicial system, making judgments difficult in lawsuits against data leakage caused by differentially private analysis.

Blockchain technology is increasingly regarded as the future of medical data storage and sharing due to its secure and tamper-proof nature. By utilizing a distributed, secured, and shared ledger, blockchain records data in a structured manner. Integrating blockchain with differential privacy offers a compelling solution, ensuring secure patient data storage while preserving privacy during data analysis and research.

\section*{Methods}
\label{section:Methods}
We followed the steps shown in Figure~\ref{queryrelease-steps} to publish our reports as explained below:

 \textbf{1. Selecting the Queries: }For our case study, we want to publish a differentially private report on the effect of arrhythmias on the QRS duration (the length of the QRS complex in milliseconds, see Figure~\ref{fig1:ECG waves}). In other words, we want to share the mean QRS duration values for different heart conditions such as atrial fibrillation and sinus bradycardia. We have ten types of arrhythmias in our dataset plus the sinus healthy rhythm, therefore we will publish eleven numbers. The normal QRS duration in adults is usually between 80 to 100 ms (children might have shorter QRS duration). The QRS duration is an important index in cardiology and is related to many heart diseases. The QRS duration becomes longer when electrical activity takes a longer time to travel throughout the ventricular myocardium. For instance, a prolonged QRS duration is linked to many conditions such as bundle branch block or congestive heart failure ~\cite{Ravietal2006}. Researchers also found a link between a prolonged QRS duration and an increased risk of death or hospitalization among patients with atrial fibrillation ~\cite{Whitbecketal2013}. In addition to mean values, we will publish the differentially private median values of QRS duration for different conditions. The median is considered a better statistic because it is not affected by outliers and might have less sensitivity. Additionally, we will publish a second report that contains the histogram or distribution (count queries) of different variables in the dataset. We will publish fourteen histograms for different variables in our dataset such as age, sex and QRS duration. 

\textbf{2. Select the Type of Differential Privacy: }Considering the sensitivity of medical data, we choose pure $\epsilon$ differential privacy to have a full guarantee of the privacy loss for our ECG analysis case.

\textbf{3. Estimate the Global Sensitivity of the Query Functions: } Many differential privacy tools use the concept of bounded contribution to simplify the calculations. In this setting, it is assumed that each user contribution has been aggregated to a single record. Additionally, in bounded contribution systems, operations such as the mean or summation require lower and upper bounds for the variables to be specified. Values smaller or greater than the lower/upper bound are clamped to limit the contribution of each person. Usually, bounds are set based on  some prior knowledge or based on local data. For instance, consider that we want to publish a differentially private number of the average age of subjects in our dataset. We know from our dataset that everyone is between 4-98 years. However, we should use the global range for this attribute. For example, we could set the upper bound for age to 100 since as of 2019, less than 0.005\% of the world's population is aged over 100; or to be completely safe, we can set the upper bound to 120. For our case study, we simply used the range of QRS duration from our dataset, which is 18-256 ms which was quite large. This means that QRS duration values smaller than 18 or larger than 256 will be clamped to 18 or 256, respectively. In this case, we used the local sensitivity for QRS duration from our dataset as the global sensitivity. We made this choice based on our empirical knowledge of QRS duration.

\textbf{Choosing $\epsilon$: }We applied the economic method proposed by Hsu et al.~\cite{Hsu} (explained earlier) to find the range of feasible values for $\epsilon$ in our analysis. To calculate participants’ belief about the cost if they do not participate in a study, we used the general statistics and costs related to health data breaches in the United States. The average cost of a health care breach in the United States in 2019 was reported to be \$429, which was published by the Ponemon Institute/IBM Security on its 2019 Cost of a Data Breach Report. The chances of an individual being affected by a medical data breach have been relatively high in the past decade. On average, each year, 23.5 million American residents have been affected by a health data breach in the past ten years. We divided this number by the total population to estimate the average chance of medical data leakage per person in the United States (approximately 8\%). Therefore, we can calculate the base cost to be $E=0.08\times429\approx34$ (expected monetary loss for each individual due to medical record leakage from another source outside of our experiment). We assume that we have \$10,000 budget (B) for this experiment and there are 10,646 (N) participants in our dataset. Plugging these values into Hsu et al.'s equation(4) ($(e^{\epsilon}-1)EN \leq B)$)~\cite{Hsu} tells us that this experiment is feasible for epsilon values below 0.027.  However, we realized that this $\epsilon$ value is too low to publish eleven queries (the mean QRS duration per condition) for the mean and median reports or fourteen count queries for the histogram report with acceptable accuracy, and the results were utterly useless due to the large amount of noise added. Instead, we were forced to adopt a higher value of 0.2 empirically (based on the previous research in this field) to generate useful reports.

\textbf{Distribute $\epsilon$ Across All the Queries: }A simple rule to manually distribute the privacy budget among some queries is to break down the budget based on the priority of queries in terms of utility. We assign more budget to queries for which we need more accurate results. To keep it simple, for our ECG Query release example, we assigned it equally across all arrhythmia types so that $\frac{0.2}{11}$ privacy budget is allocated for each mean/median query and $\frac{0.2}{14}$ for each histogram. A better approach would be to use a weighted approach where we allocate more budget to the less common conditions to balance the utility across different arrhythmias.

\textbf{Publish the Results: } 
Finally, we run the queries based on the chosen parameters and publish the results. We report $\epsilon$ with each report.

\section*{Data Availability}
The data that support the findings of this study are openly available at~\cite{zheng2020} \\
\url{https://figshare.com/collections/ChapmanECG/4560497}.

\section*{Code Availability}
The Code for the analysis used in this paper is openly available at\\ \url{https://github.com/arin-gzn/differential-privacy-arrhythmia-analysis}


\begin{thebibliography}{10}

\bibitem{ZhengJ2020}
J.~Zheng, H.~Chu, D.~Struppa, J.~Zhang, S.~M. Yacoub, H.~El-Askary, A.~Chang, L.~Ehwerhemuepha, I.~Abudayyeh, A.~Barrett, G.~Fu, H.~Yao, D.~Li, H.~Guo, and C.~Rakovski.
\newblock Optimal multi-stage arrhythmia classification approach.
\newblock {\em Sci Rep}, 10(1):2898, 2020.

\bibitem{Zheng2021}
Jianwei {Zheng}, Guohua {Fu}, Islam {Abudayyeh}, Magdi {Yacoub}, Anthony {Chang}, William~W. {Feaster}, Louis {Ehwerhemuepha}, Hesham {El-Askary}, Xianfeng {Du}, Bin {He}, Mingjun {Feng}, Yibo {Yu}, Binhao {Wang}, Jing {Liu}, Hai {Yao}, Huimin {Chu}, and Cyril {Rakovski}.
\newblock A high precision machine learning algorithm to classify left and right outflow tract ventricular tachycardia.
\newblock {\em Frontiers in Psychology}, s, 2021.

\bibitem{Zhengjw2020}
Jianwei Zheng, Guohua Fu, Kyle Anderson, Huimin Chu, and Cyril Rakovski.
\newblock A 12-lead ecg database to identify origins of idiopathic ventricular arrhythmia containing 334 patients.
\newblock {\em Scientific Data}, 7(1):98, 2020.

\bibitem{Biel}
L.~{Biel}, O.~{Pettersson}, L.~{Philipson}, and P.~{Wide}.
\newblock Ecg analysis: a new approach in human identification.
\newblock {\em IEEE Transactions on Instrumentation and Measurement}, 50(3):808--812, 2001.

\bibitem{Carreiras}
C.~{Carreiras}, A.~{Lourenço}, A.~{Fred}, and R.~{Ferreira}.
\newblock Ecg signals for biometric applications - are we there yet?
\newblock In {\em 2014 11th International Conference on Informatics in Control, Automation and Robotics (ICINCO)}, volume~02, pages 765--772, 2014.

\bibitem{Attia2019AgeAS}
Zachi~I. Attia, P.~Friedman, P.~Noseworthy, F.~L{\'o}pez-Jim{\'e}nez, Dorothy~J. Ladewig, Gaurav Satam, P.~A. Pellikka, T.~Munger, S.~Asirvatham, C.~Scott, R.~Carter, and S.~Kapa.
\newblock Age and sex estimation using artificial intelligence from standard 12-lead ecgs.
\newblock {\em Circulation: Arrhythmia and Electrophysiology}, 12:e007284 -- e007284, 2019.

\bibitem{Noseworthy}
Attia Z. I. Brewer L. C. Hayes S. N. Yao X. Kapa S. Friedman P. A. Lopez-Jimenez~F. Noseworthy, P.~A.
\newblock Assessing and mitigating bias in medical artificial intelligence: The effects of race and ethnicity on a deep learning model for ecg analysis. circulation.
\newblock {\em Arrhythmia and electrophysiology}, 13 3, 2020.

\bibitem{Cabra2018WideML}
Jose-Luis Cabra, D.~Mendez, and Luis~C. Trujillo.
\newblock Wide machine learning algorithms evaluation applied to ecg authentication and gender recognition.
\newblock In {\em ICBEA '18}, 2018.

\bibitem{Wiggins2008EvolvingAB}
Matthew~C. Wiggins, A.~Saad, B.~Litt, and G.~Vachtsevanos.
\newblock Evolving a bayesian classifier for ecg-based age classification in medical applications.
\newblock {\em Applied soft computing}, 8 1:599--608, 2008.

\bibitem{Koelstra}
S.~{Koelstra}, C.~{Muhl}, M.~{Soleymani}, J.~{Lee}, A.~{Yazdani}, T.~{Ebrahimi}, T.~{Pun}, A.~{Nijholt}, and I.~{Patras}.
\newblock Deap: A database for emotion analysis ;using physiological signals.
\newblock {\em IEEE Transactions on Affective Computing}, 3(1):18--31, 2012.

\bibitem{BRAS2018}
Susana Brás, Jacqueline H.~T. Ferreira, Sandra~C. Soares, and Armando~J. Pinho.
\newblock Biometric and emotion identification an ecg compression based method.
\newblock {\em Frontiers in Psychology}, 9:467, 2018.

\bibitem{Sweeney2}
Latanya Sweeney.
\newblock Simple demographics often identify people uniquely.
\newblock 2000.

\bibitem{Hayes}
B.~Hayes.
\newblock Uniquely me.
\newblock {\em American Scientist}, 102:106--109, 2014.

\bibitem{Sweeney}
L.~Sweeney.
\newblock Only you, your doctor and many others may know.
\newblock {\em Technology Science}, 2015.

\bibitem{Liuetal2018}
Yongtai Liu, Zhiyu Wan, Weiyi Xia, Murat Kantarcioglu, Yevgeniy Vorobeychik, Ellen~Wright Clayton, Abel Kho, David Carrell, and Bradley~A Malin.
\newblock Detecting the presence of an individual in phenotypic summary data.
\newblock {\em AMIA ... Annual Symposium proceedings. AMIA Symposium}, 2018:760—769, 2018.

\bibitem{Dwork}
C.~Dwork and A.~Roth.
\newblock {\em The Algorithmic Foundations of Differential Privacy}.
\newblock 2014.

\bibitem{DinurNissim}
Irit Dinur and Kobbi Nissim.
\newblock Revealing information while preserving privacy.
\newblock PODS '03, page 202–210, New York, NY, USA, 2003. Association for Computing Machinery.

\bibitem{JayaramanEvans}
Bargav Jayaraman and David Evans.
\newblock Evaluating differentially private machine learning in practice.
\newblock In {\em 28th {USENIX} Security Symposium ({USENIX} Security 19)}, pages 1895--1912, Santa Clara, CA, August 2019. {USENIX} Association.

\bibitem{pmlr-v70-rubinstein17a}
Benjamin I.~P. Rubinstein and Francesco Ald{\`a}.
\newblock Pain-free random differential privacy with sensitivity sampling.
\newblock In Doina Precup and Yee~Whye Teh, editors, {\em Proceedings of the 34th International Conference on Machine Learning}, volume~70 of {\em Proceedings of Machine Learning Research}, pages 2950--2959. PMLR, 06--11 Aug 2017.

\bibitem{Bernstein}
Francesco Ald\`{a} and Benjamin~I.P. Rubinstein.
\newblock The bernstein mechanism: Function release under differential privacy.
\newblock In {\em Proceedings of the Thirty-First AAAI Conference on Artificial Intelligence}, AAAI'17, page 1705–1711. AAAI Press, 2017.

\bibitem{Maetal2023}
Chuan Ma, Jun Li, Ming Ding, Bo~Liu, Kang Wei, Jian Weng, and H.~Vincent Poor.
\newblock Rdp-gan: A rényi-differential privacy based generative adversarial network.
\newblock {\em IEEE Transactions on Dependable and Secure Computing}, 20(6):4838--4852, 2023.

\bibitem{Huangetal2019}
Pei Huang, Linke Guo, Ming Li, and Yuguang Fang.
\newblock Practical privacy-preserving ecg-based authentication for iot-based healthcare.
\newblock {\em IEEE Internet of Things Journal}, 6(5):9200--9210, 2019.

\bibitem{Habibaetal2021}
Mansura Habiba, Eoin Borphy, Barak~A. Pearlmutter, and Tomas Ward.
\newblock Ecg synthesis with neural ode and gan models.
\newblock In {\em 2021 International Conference on Electrical, Computer and Energy Technologies (ICECET)}, pages 1--6, 2021.

\bibitem{Jafarlouetal2022}
Salar Jafarlou, Amir~M. Rahmani, Nikil Dutt, and Sanaz~Rahimi Mousavi.
\newblock Ecg biosignal deidentification using conditional generative adversarial networks.
\newblock In {\em 2022 44th Annual International Conference of the IEEE Engineering in Medicine \& Biology Society (EMBC)}, pages 1366--1370, 2022.

\bibitem{YING2023101568}
Zuobin Ying, Guoyang Zhang, Zijie Pan, Chiawei Chu, and Ximeng Liu.
\newblock Fedecg: A federated semi-supervised learning framework for electrocardiogram abnormalities prediction.
\newblock {\em Journal of King Saud University - Computer and Information Sciences}, 35(6):101568, 2023.

\bibitem{Wangetal2015}
Zhiguang Wang and Tim Oates.
\newblock Imaging time-series to improve classification and imputation.
\newblock 05 2015.

\bibitem{Liuetal2020}
Xin Liu, Pan Zhou, Tie Qiu, and Dapeng~Oliver Wu.
\newblock Blockchain-enabled contextual online learning under local differential privacy for coronary heart disease diagnosis in mobile edge computing.
\newblock {\em IEEE Journal of Biomedical and Health Informatics}, 24(8):2177--2188, 2020.

\bibitem{agrawal2024federated}
Vikhyat Agrawal, Sunil~Vasu Kalmady, Venkataseetharam~Manoj Malipeddi, Manisimha~Varma Manthena, Weijie Sun, Saiful Islam, Abram Hindle, Padma Kaul, and Russell Greiner.
\newblock Federated learning and differential privacy techniques on multi-hospital population-scale electrocardiogram data.
\newblock {\em arXiv preprint arXiv:2405.00725}, 2024.

\bibitem{Sonetal}
{Son et al}.
\newblock Privacy-preserving electrocardiogram monitoring for intelligent arrhythmia detection.
\newblock {\em Sensors (Basel)}, 17(6):1360, 2017.

\bibitem{Bonomietal2022}
Luca Bonomi, Zeyun Wu, and Liyue Fan.
\newblock {Sharing personal ECG time-series data privately}.
\newblock {\em Journal of the American Medical Informatics Association}, 29(7):1152--1160, 04 2022.

\bibitem{Chatzikokolakis2013BroadeningTS}
Konstantinos Chatzikokolakis, Miguel~E. Andr{\'e}s, Nicol{\'a}s~Emilio Bordenabe, and Catuscia Palamidessi.
\newblock Broadening the scope of differential privacy using metrics.
\newblock In {\em International Symposium on Privacy Enhancing Technologies}, 2013.

\bibitem{BalazsDesfontaines}
Balázs Pejó and Damien Desfontaines.
\newblock Sok: Differential privacies.
\newblock 2020.

\bibitem{Hall_Wasserman_Rinaldo_2013}
Robert Hall, Larry Wasserman, and Alessandro Rinaldo.
\newblock Random differential privacy.
\newblock {\em Journal of Privacy and Confidentiality}, 4(2), Mar. 2013.

\bibitem{barber2014privacy}
Rina~Foygel Barber and John~C. Duchi.
\newblock Privacy and statistical risk: Formalisms and minimax bounds, 2014.

\bibitem{CuffYu}
Paul Cuff and Lanqing Yu.
\newblock Differential privacy as a mutual information constraint.
\newblock In {\em Proceedings of the 2016 ACM SIGSAC Conference on Computer and Communications Security}, CCS '16, page 43–54, New York, NY, USA, 2016. Association for Computing Machinery.

\bibitem{Ilyaetal2017}
Ilya Mironov.
\newblock Rényi differential privacy.
\newblock In {\em 2017 IEEE 30th Computer Security Foundations Symposium (CSF)}, pages 263--275, 2017.

\bibitem{NissimRaskhodnikova}
Kobbi Nissim, Sofya Raskhodnikova, and Adam Smith.
\newblock Smooth sensitivity and sampling in private data analysis.
\newblock In {\em STOC'07}, Proceedings of the Annual ACM Symposium on Theory of Computing, pages 75--84, 2007.
\newblock Copyright: Copyright 2011 Elsevier B.V., All rights reserved.; STOC'07: 39th Annual ACM Symposium on Theory of Computing ; Conference date: 11-06-2007 Through 13-06-2007.

\bibitem{near_abuah_2021}
Joseph~P. Near and Chiké Abuah.
\newblock {\em Programming Differential Privacy}, volume~1.
\newblock 2021.

\bibitem{Lee2011HowMI}
Jaewoo Lee and C.~Clifton.
\newblock How much is enough? choosing $\epsilon$ for differential privacy.
\newblock In {\em ISC}, 2011.

\bibitem{Naldi2015DifferentialPA}
M.~Naldi and G.~D'Acquisto.
\newblock Differential privacy: An estimation theory-based method for choosing epsilon.
\newblock {\em ArXiv}, abs/1510.00917, 2015.

\bibitem{Kohli}
N.~{Kohli} and P.~{Laskowski}.
\newblock Epsilon voting: Mechanism design for parameter selection in differential privacy.
\newblock In {\em 2018 IEEE Symposium on Privacy-Aware Computing (PAC)}, pages 19--30, 2018.

\bibitem{Greenberg}
Andy Greenberg.
\newblock How one of apple's key privacy safeguards falls short.
\newblock {\em Wired}, 2017.

\bibitem{Erlingsson}
\'{U}lfar Erlingsson, Vasyl Pihur, and Aleksandra Korolova.
\newblock Rappor: Randomized aggregatable privacy-preserving ordinal response.
\newblock In {\em Proceedings of the 2014 ACM SIGSAC Conference on Computer and Communications Security}, CCS '14, page 1054–1067, New York, NY, USA, 2014. Association for Computing Machinery.

\bibitem{Dingetal}
Bolin {Ding}, Janardhan {Kulkarni}, and Sergey {Yekhanin}.
\newblock {Collecting Telemetry Data Privately}.
\newblock {\em arXiv e-prints}, page arXiv:1712.01524, December 2017.

\bibitem{Korolova2009ReleasingSQ}
Aleksandra Korolova, K.~Kenthapadi, Nina Mishra, and A.~Ntoulas.
\newblock Releasing search queries and clicks privately.
\newblock In {\em WWW '09}, 2009.

\bibitem{Machanavajjhalaetal}
Ashwin Machanavajjhala, Aleksandra Korolova, and Atish~Das Sarma.
\newblock Personalized social recommendations: Accurate or private.
\newblock {\em Proc. VLDB Endow.}, 4(7):440–450, April 2011.

\bibitem{Cormodeetal}
Graham Cormode, Magda Procopiuc, Entong Shen, Divesh Srivastava, and Ting Yu.
\newblock Differentially private spatial decompositions.
\newblock {\em CoRR}, abs/1103.5170, 2011.

\bibitem{Acs-Castellucia}
Gergely {\'{A}}cs and Claude Castelluccia.
\newblock {DREAM:} differentially private smart metering.
\newblock {\em CoRR}, abs/1201.2531, 2012.

\bibitem{Bhaskaretal}
Raghav Bhaskar, Srivatsan Laxman, Adam Smith, and Abhradeep Thakurta.
\newblock Discovering frequent patterns in sensitive data.
\newblock In {\em Proceedings of the 16th ACM SIGKDD International Conference on Knowledge Discovery and Data Mining}, KDD '10, page 503–512, New York, NY, USA, 2010. Association for Computing Machinery.

\bibitem{Uhler_Slavkovic_Fienberg_2013}
Caroline Uhler, Aleksandra~B. Slavkovic, and Stephen~E. Fienberg.
\newblock Privacy-preserving data sharing for genome-wide association studies.
\newblock {\em Journal of Privacy and Confidentiality}, 5(1), Aug. 2013.

\bibitem{Hsu}
Justin Hsu, Marco Gaboardi, Andreas Haeberlen, Sanjeev Khanna, Arjun Narayan, {Benjamin C.} Pierce, and Aaron Roth.
\newblock Differential privacy: an economic method for choosing epsilon.
\newblock In {\em Proceedings of the 2014 IEEE 27th Computer Security Foundations Symposium, CSF 2014}, pages 398--410. IEEE Computer Society, 2014.

\bibitem{LiHay}
Chao Li, Michael Hay, Vibhor Rastogi, Gerome Miklau, and Andrew McGregor.
\newblock Optimizing linear counting queries under differential privacy.
\newblock In {\em Proceedings of the Twenty-Ninth ACM SIGMOD-SIGACT-SIGART Symposium on Principles of Database Systems}, PODS '10, page 123–134, New York, NY, USA, 2010. Association for Computing Machinery.

\bibitem{Hardtetal}
Moritz and Kunal Talwar.
\newblock On the geometry of differential privacy.
\newblock In {\em Proceedings of the Forty-Second ACM Symposium on Theory of Computing}, STOC '10, page 705–714, New York, NY, USA, 2010. Association for Computing Machinery.

\bibitem{DworkNaReRoVa09}
Cynthia Dwork, Moni Naor, Omer Reingold, Guy Rothblum, and Salil Vadhan.
\newblock On the complexity of differentially private data release: Efficient algorithms and hardness results.
\newblock In {\em Proceedings of the 41st Annual ACM Symposium on Theory of Computing (STOC {\textquoteleft}09)}, page 381{\textendash}390, Bethesda, MD, 31 May{\textendash}2 June 2009.

\bibitem{zheng2020}
Jianwei Zheng, Jianming Zhang, Sidy Danioko, Hai Yao, Hangyuan Guo, and Cyril Rakovski.
\newblock A 12-lead electrocardiogram database for arrhythmia research covering more than 10,000 patients.
\newblock {\em Scientific Data}, 7, 02 2020.

\bibitem{Tran2023FairDPCF}
Khang~Hoang Tran, Ferdinando Fioretto, Issa Khalil, My~T. Thai, and Nhathai Phan.
\newblock Fairdp: Certified fairness with differential privacy.
\newblock {\em ArXiv}, abs/2305.16474, 2023.

\bibitem{Jagielski2018DifferentiallyPF}
Matthew Jagielski, Michael Kearns, Jieming Mao, Alina Oprea, Aaron Roth, Saeed Sharifi-Malvajerdi, and Jonathan Ullman.
\newblock Differentially private fair learning.
\newblock {\em ArXiv}, abs/1812.02696, 2018.

\bibitem{Hardt2016EqualityOO}
Moritz Hardt, Eric Price, and Nathan Srebro.
\newblock Equality of opportunity in supervised learning.
\newblock {\em ArXiv}, abs/1610.02413, 2016.

\bibitem{biswas2021machine}
Sreemoyee Biswas, Nilay Khare, Pragati Agrawal, and Priyank Jain.
\newblock Machine learning concepts for correlated big data privacy.
\newblock {\em Journal of Big Data}, 8(1):1--32, 2021.

\bibitem{gehrke2011towards}
Johannes Gehrke, Edward Lui, and Rafael Pass.
\newblock Towards privacy for social networks: A zero-knowledge based definition of privacy.
\newblock In {\em Theory of cryptography conference}, pages 432--449. Springer, 2011.

\bibitem{Santos-Lozada13405}
Alexis~R. Santos-Lozada, Jeffrey~T. Howard, and Ashton~M. Verdery.
\newblock How differential privacy will affect our understanding of health disparities in the united states.
\newblock {\em Proceedings of the National Academy of Sciences}, 117(24):13405--13412, 2020.

\bibitem{WezerekRiper2020}
Gus Wezerek and David~Van Riper.
\newblock Changes to the census could make small towns disappear.
\newblock {\em The New York Times}, 2 2020.

\bibitem{Ravietal2006}
Thomas J. Wang Byung-Ho Nam Emelia J. Benjamin Daniel Levy Martin G. Larson William B. Kannel Ralph B. D’Agostino Ramachandran S.~Vasan Ravi~Dhingra, Michael J.~Pencina.
\newblock Electrocardiographic qrs duration and the risk of congestive heart failure.
\newblock {\em Hypertension}, 47(5):861--867, 2006.

\bibitem{Whitbecketal2013}
Matthew~G. Whitbeck, Richard~J. Charnigo, Jignesh Shah, Gustavo Morales, Steve~W. Leung, Brandon Fornwalt, Alison~L. Bailey, Khaled Ziada, Vincent~L. Sorrell, Milagros~M. Zegarra, Jenks Thompson, Neil~Aboul Hosn, Charles~L. Campbell, John Gurley, Paul Anaya, David~C. Booth, Luigi~Di Biase, Andrea Natale, Susan Smyth, David~J. Moliterno, Claude~S. Elayi, and the AFFIRM~investigators.
\newblock {QRS duration predicts death and hospitalization among patients with atrial fibrillation irrespective of heart failure: evidence from the AFFIRM study}.
\newblock {\em EP Europace}, 16(6):803--811, 12 2013.

\end{thebibliography}

\section*{Author Contributions}

This study was part of the PhD research of Arin Ghazarian under the supervision of Cyril Rakovski. Arin Ghazarian has drafted the manuscript. Jianwei Zheng has acquired and prepared the data. All authors have reviewed and provided feedback on the manuscript.
\section*{Competing Interests}
The authors declare no competing interests.
\end{document}